\documentclass{ptapap}

\author{Pawe{\l} Pietrukowicz}[Warsaw]
\affil[Warsaw]{Warsaw University Observatory\\
Al. Ujazdowskie 4, 00--478 Warszawa, Poland}

\headtitle{Properties of BLAPs}
		
\title{On the Properties of Blue Large-Amplitude Pulsators.
No BLAPs in the Magellanic Clouds}

\begin{document}

\maketitle

\begin{abstract}
We present the properties of the recently discovered class of variable
stars, Blue Large-Amplitude Pulsators (BLAPs). These extremely rare,
short-period pulsating objects were detected thanks to regular,
high-cadence observations of hundreds of millions of Milky Way
stars by the OGLE variability survey. The new variables closely
resemble classical pulsators, Cepheids, and RR~Lyrae-type stars, but at
effective temperatures at which pulsations are due to the presence
of iron-group elements. Theory shows that BLAPs are evolved low-mass
stars with a giant-like structure, but their origin remains a mystery.
In this contribution, we report the negative result of a search
for BLAPs in the whole Magellanic System.
\end{abstract}

Detection of the first variable showing regular, Cepheid-like brightness
variations with an exceptionally large $I$-band amplitude of $0.24$\,mag
at the very short period of $28.26$\,min was a serendipitous result of
searches for periodic variables in Galactic disc fields monitored by
the Optical Gravitational Lensing Experiment \citep[OGLE]{2013AcA....63..379P}.
OGLE is a long-term, wide-field variability survey conducted with the
$1.3$\,m Warsaw telescope at Las Campanas Observatory, Chile. It started
in 1992 with the original aim of detecting microlensing events towards the
Galactic bulge \citep{1992AcA....42..253U}. Since then, OGLE has detected
and classified about one million new variable stars
\citep[e.g.,][]{2013AcA....63...21S,2014AcA....64..177S,2016AcA....66..405S,2015AcA....65..313M}.
In its current fourth phase, OGLE-IV monitors about one billion stars
over $3500$\,deg$^2$ of the Galactic bulge, Galactic disc, and Magellanic
System \citep{2015AcA....65....1U}.

The unusual short-period variable was tentatively classified as a
$\delta$~Scuti-type pulsator despite the amplitude is several times higher
than amplitudes observed among the shortest-period $\delta$~Sct stars.
Definitive classification required follow-up observations.
Low-resolution spectroscopy showed that this object is much hotter
than $\delta$~Sct stars, and its surface gravity is higher than
in main sequence stars \citep{2015AcA....65...63P}. The proof for the
pulsation nature of this mysterious object was provided with
moderate-resolution spectra taken at opposite phases of the variability
cycle \citep{2017NatAs...1E.166P}. By fitting model atmospheres to the
combined spectrum, the following parameters were obtained: effective
temperature $T_{\rm eff}=30800\pm500$\,K, surface gravity
$\log{g}=4.61\pm0.07$, helium-to-hydrogen ratio
$\log{(N_{\rm He}/N_{\rm H})}=-0.55\pm0.05$. True mean values can be
slightly different due to difficulties in observations of such rapidly
varying and relatively faint object ($\langle V \rangle = 17.71$\,mag,
$\langle I \rangle = 17.22$\,mag).

Another thirteen variables with very similar photometric behaviour were found in
the OGLE-IV Galactic bulge fields. All variables have exceptionally large
amplitudes of about $0.2-0.4$\,mag in $V$ and $I$ bands at very short periods
of roughly $20-40$\,min. Colour-magnitude diagrams constructed for observed
fields with the variables show that these stars are located far blueward of
the main sequence. The variables have mean $I$-band brightness
between $16.7$ and $18.5$\,mag but one brighter object of $I\approx15.1$\,mag.
The bulge variables are moderately reddened if one compares their observed
colour $V-I$ between $+0.1$ and $+1.3$\,mag with the expected intrinsic value of
$(V-I)_0=-0.29$\,mag. Spectroscopic data for three of the bulge stars confirmed
that their atmosphere parameters are practically identical to the ones
obtained for the Galactic disc object. All variables seem to form
a homogeneous class of stars. Based on the observed properties, the name
Blue Large-Amplitude Pulsators (BLAPs) was proposed \citep{2017NatAs...1E.166P}.
The first detected pulsator, OGLE-BLAP-001, can be treated as the prototype
of the whole class. Details on each star are provided in Table~\ref{tab}.

Such photometric variations are not seen in any other hot pulsators,
i.e., stars in which pulsations are driven by the $\kappa$-mechanism due to
the presence of a metal bump in the opacity curve. Envelope models presented
in \citet{2017NatAs...1E.166P} demonstrate that BLAPs have a giant-like
structure. Due to their inflated envelopes, they cannot be treated as dwarfs
or subdwarfs. At similar effective temperatures, BLAPs are at least by
an order of magnitude more luminous than hot subdwarf B-type (sdB) stars
\citep{2016PASP..128h2001H}. Long-term OGLE observations show that light
curves of BLAPs are very stable over time. Period change rates of the order of
$10^{-7}$\,yr$^{-1}$ indicate that these stars evolve on the nuclear time scale.

There are, however, many fundamental questions related to the new stars:
what is their exact mass and luminosity, what is their real metallicity, are
they single objects, how did they form? Linking the envelope models developed
for the prototype object with available stellar evolutionary models leads
to two configurations of different masses. BLAPs are either
$\sim$1.0\,M$_{\odot}$ stars with a helium-burning core or $\sim$0.3\,M$_{\odot}$
stars with a hydrogen-burning shell above a degenerate helium core.
None of these configurations can be produced in the evolution of
a single star. The fact that very few BLAPs are known points to a rare
episode in stellar evolution. They can be remnants of stellar mergers.
However, no signs of binarity in the obtained data supports this hypothesis.

The luminosity of the new pulsators was derived from envelope models.
For the prototype object, depending on the configuration:
${\rm log}L/{\rm L}_{\odot}=2.3$ in the case of the $\sim$0.3\,M$_{\odot}$ star,
${\rm log}L/{\rm L}_{\odot}=2.6$ in the case of the $\sim$1.0\,M$_{\odot}$ star.
These luminosities can be transformed to the following $V$ and $I$-band
absolute magnitudes: $M_V=+1.95$ and $M_I=+2.24$ in the less massive case,
$M_V=+1.20$ and $M_I=+1.49$ in the more massive one.

Accurate determination of the mass and luminosity would be possible
for a pulsator located at known distance. Discovery of such
a star in the Large Magellanic Cloud (LMC) or Small Magellanic Cloud
(SMC) would be very helpful. The expected brightness range of BLAPs at the
LMC's distance \citep[$\approx$50.0\,kpc,][]{2013Natur.495...76P}
is $19.20<V<20.45$\,mag and $19.49<I<20.74$\,mag, assuming negligible extinction.
At the SMC's distance \citep[$\approx$62.1\,kpc,][]{2014ApJ...780...59G},
BLAPs would have $19.49<V<20.74$\,mag and $19.94<I<21.19$\,mag.

We have searched for BLAPs in the rich OGLE-IV data collected for the
Magellanic System in years $2010-2015$. The monitored area consists of
475 OGLE-IV fields covering a total of about 650\,deg$^2$. The number
of $I$-band measurements varies from field to field. There are $\sim$700
frames collected for the LMC, $\sim$600 frames for the SMC, and $\sim$400 frames
for the Magellanic Bridge area. Standard OGLE observations, taken with
exposure times of 150\,s, reach $V\approx22.5$\,mag and $I\approx22$\,mag,
but they are complete down to $V\approx21$\,mag and $I\approx20.5$\,mag.

We searched 77.9 million stellar detections for periodic
signals in the frequency range of $30-100$ cycles per day
or the period range of $14.4-48$\,min, with the FNPEAKS
code\footnote{http://helas.astro.uni.wroc.pl/deliverables.php?lang=en\&active=fnpeaks}.
Light curves of $46\,830$ stars with a signal-to-noise ratio $>5$ were
subject to visual inspection. Unfortunately, we did not find any BLAP.
In the analyzed period range, we detected only one {\it bona fide}
variable, a foreground SX~Phoenicis-type (or population II $\delta$~Sct-type)
star LMC512.02.16620 at equatorial coordinates
(RA,Dec)$_{2000.0}=(05$:10:25.55, $-66$:37:55.7), with mean brightness
$I=16.74$\,mag, colour $V-I=0.30$\,mag, period of $31.02$\,min, and $I$-band
amplitude of $0.013$\,mag. Our inspection led to the detection of four new
short-period eclipsing binary stars, LMC527.10.3916, LMC556.21.1214,
LMC570.22.45, and SMC732.19.320, with orbital periods between $1.68$ and $3.36$\,h
and three faint unknown $\delta$~Sct stars, LMC508.17.28064, SMC731.04.3587,
and LMC570.26.6352, with pulsation periods of 81.25, 82.18, 91.69\,min,
mean $I$-band brightness of 20.09, 19.79, 20.57\,mag, and amplitudes of
0.32, 0.35, 0.41\,mag, respectively.

The result of our search for BLAPs in the Magellanic System is negative.
However, we cannot exclude a possibility that some low-luminosity variables
of this type reside in the Magellanic Clouds. In our calculations, we did not
take interstellar extinction into account. Detection of stellar sources
from the OGLE standard images is about 95\% complete at $I=20.7$\,mag
and about 80\% at $I=21.2$\,mag. Completeness of our search for variables
is likely lower, but the detection of faint $\delta$~Sct stars demonstrates
that some large amplitude variables can be found. From the observational
point of view, chances for BLAPs seem to be higher in the more distant SMC.
On the other hand, the SMC is less metal-rich than the LMC and much less than the Milky Way.
Recent theoretical models \citep[][A. Pamyatnykh, priv. comm.]{2016MNRAS.458.1352J}
show that pulsation instability in hydrogen-deficient atmospheres appears
at high (solar) metallicity in the region occupied by BLAPs on the
$T_{\rm eff}-\log{g}$ plane. The lack of pulsations at low
metallicity is consistent with the fact that BLAPs are not observed
in the Galactic halo and globular clusters.

Distance measurements to known BLAPs from the \emph{Gaia} satellite should
help us determine accurate luminosities and masses of these objects
and to find out what their true structure is. Future high-resolution
spectra should provide information on the exact iron content and
presence of other elements in the atmospheres of BLAPs.

\begin{table}
\caption{Parameters of known Blue Large-Amplitude Pulsators}
\label{tab}
\begin{center}
\rotatebox{90}{
{\scriptsize
\begin{tabular}{llllllllllll}
\hline
\hline
Variable name & RA(2000.0) & Dec(2000.0) & $\langle V \rangle$ & $\langle I \rangle$ & amp$_V$ & amp$_I$ & $P$ & $r$ & $T_{\rm eff}$ & ${\rm log}g$ & log($N_{\rm He}$/$N_{\rm H}$)\\
 & [h:m:s] & [d:m:s] & [mag] & [mag] & [mag] & [mag] & [min] & [$10^{-7}$\,y$^{-1}$] & [K] & [cm s$^{-2}$] & \\
\hline
OGLE-BLAP-001 & 10:41:48.77 & $-61$:25:08.5 & 17.71 & 17.22 & 0.41 & 0.24 & 28.255 & $+2.9\pm3.7$   & $30800\pm500$ & $4.61\pm0.07$ & $-0.55\pm0.05$ \\
OGLE-BLAP-002 & 17:43:58.02 & $-19$:16:54.1 & 18.89 & 17.95 & 0.32 & 0.36 & 23.286 &                & & & \\
OGLE-BLAP-003 & 17:44:51.48 & $-24$:10:04.0 & 19.19 & 18.10 & 0.27 & 0.23 & 28.458 & $+0.82\pm0.32$ & & & \\
OGLE-BLAP-004 & 17:51:04.72 & $-22$:09:03.4 & 18.81 & 17.53 & 0.43 & 0.26 & 22.357 &                & & & \\
OGLE-BLAP-005 & 17:52:18.73 & $-31$:56:35.0 & 19.99 & 18.77 & 0.33 & 0.30 & 27.253 & $+0.63\pm0.26$ & & & \\
OGLE-BLAP-006 & 17:55:02.88 & $-29$:50:37.5 & 18.35 & 17.38 & 0.28 & 0.23 & 38.015 & $-2.85\pm0.31$ & & & \\
OGLE-BLAP-007 & 17:55:57.52 & $-28$:52:11.0 & 19.27 & 18.46 & 0.33 & 0.28 & 35.182 & $-2.40\pm0.51$ & & & \\
OGLE-BLAP-008 & 17:56:48.26 & $-32$:21:35.6 & 18.73 & 17.59 & 0.28 & 0.19 & 34.481 & $+2.11\pm0.27$ & & & \\
OGLE-BLAP-009 & 17:58:48.20 & $-27$:16:53.7 & 15.65 & 15.07 & 0.29 & 0.24 & 31.935 & $+1.63\pm0.08$ & $31800\pm1400$ & $4.40\pm0.18$ & $-0.41\pm0.13$ \\
OGLE-BLAP-010 & 17:58:59.22 & $-35$:18:07.0 & 17.23 & 16.92 & 0.38 & 0.34 & 32.133 & $+0.44\pm0.21$ & & & \\
OGLE-BLAP-011 & 18:00:23.24 & $-35$:58:03.1 & 17.25 & 16.98 & 0.22 & 0.29 & 34.875 &                & $26200\pm2900$ & $4.20\pm0.29$ & $-0.45\pm0.11$ \\
OGLE-BLAP-012 & 18:05:44.20 & $-30$:11:15.2 & 18.26 & 17.62 & 0.35 & 0.34 & 30.897 & $+0.03\pm0.15$ & & & \\
OGLE-BLAP-013 & 18:05:52.70 & $-26$:48:18.0 & 19.24 & 18.19 & 0.25 & 0.22 & 39.326 & $+7.65\pm0.67$ & & & \\
OGLE-BLAP-014 & 18:12:41.79 & $-31$:12:07.8 & 16.79 & 16.68 & 0.34 & 0.26 & 33.623 & $+4.82\pm0.39$ & $30900\pm2100$ & $4.42\pm0.26$ & $-0.54\pm0.16$ \\
\hline
\end{tabular}}}
\end{center}
\end{table}

\acknowledgements{The OGLE project has received funding from the Polish
National Science Centre grant MAESTRO no. 2014/14/A/ST9/00121. P.P. is
supported by the National Science Centre grant OPUS No. 2016/23/B/ST9/00655.}

\bibliographystyle{ptapap}
\bibliography{PPietrukowicz}

\end{document}